\documentclass[onecolumn]{article}
\usepackage[letterpaper, margin=3cm]{geometry}

\usepackage[T1]{fontenc}
\usepackage{graphicx}
\usepackage{amsmath}
\usepackage{booktabs}
\usepackage{authblk}
\usepackage{xcolor}

\usepackage{hyperref}

\definecolor{blue}{HTML}{1f77b4}
\definecolor{orange}{HTML}{ff7f0e}
\definecolor{green}{HTML}{2ca02c}
\definecolor{red}{HTML}{d62728}

\usepackage{amssymb}
\usepackage{pifont}
\newcommand{\xmark}{\ding{55}}%

\usepackage[normalem]{ulem}

\begin{document}

\title{SIREN: Software Identification\\ and Recognition in HPC Systems}

\author[1]{Thomas Jakobsche}
\author[2]{Fredrik Robertsén}
\author[3]{Jessica R. Jones}
\author[4]{Utz-Uwe Haus}
\author[1]{Florina M. Ciorba}

\affil[1]{University of Basel, Switzerland\\
\texttt{\{thomas.jakobsche,florina.ciorba\}@unibas.ch}}

\affil[2]{CSC IT Center for Science, Finland\\
\texttt{fredrik.robertsen@csc.fi}}

\affil[3]{Hewlett Packard Enterprise, United Kingdom\\
\texttt{j.r.jones@hpe.com}}

\affil[4]{Hewlett Packard Enterprise, Switzerland\\
\texttt{utz-uwe.haus@hpe.com}}

\date{}

\maketitle

\begin{abstract}
    HPC systems use monitoring and operational data analytics to ensure efficiency, performance, and orderly operations. 
    Application-specific insights are crucial for analyzing the increasing complexity and diversity of HPC workloads, particularly through the \textit{identification} of unknown software and \textit{recognition} of repeated executions, which facilitate system optimization and security improvements.
    However, traditional identification methods using job or file names are unreliable for arbitrary user-provided names (\texttt{a.out}).
    Fuzzy hashing of executables detects similarities despite changes in executable version or compilation approach while preserving privacy and file integrity, overcoming these limitations.
    We introduce SIREN, a process-level data collection framework for \textbf{s}oftware \textbf{i}dentification and \textbf{re}cognitio\textbf{n}.
    SIREN improves observability in HPC by enabling analysis of process metadata, environment information, and executable fuzzy hashes.
    Findings from a first opt-in deployment campaign on LUMI show SIREN's ability to provide insights into software usage, recognition of repeated executions of known applications, and similarity-based identification of unknown applications.
\end{abstract}

\section{Introduction}

\textit{Monitoring} of High Performance Computing (HPC) systems is a well established practice that traditionally involves systematic collection of data and alerts when thresholds are exceeded.
\textit{Operational Data Analytics(ODA)}~\cite{netti2021conceptual} is the further analysis of collected data, typically relying on postmortem analysis and manual feedback/response through human-in-the-loop~\cite{brandt2023driving}.
\textit{Observability} delivers real-time application-specific insight that enables dynamic diagnosis of and actionable feedback to complex application behavior~\cite{yokelson2024soma}.
\textit{HPC Security} also motivates application-specific insight highlighted by a series of cyberattacks and reports of cryptocurrency mining on HPC systems~\cite{delwiche2014dogecoin, bbc2014agency, trader2018cryptomining, bbc2020supercomputers}.
Application-specific insights are also important for HPC user support teams, as they unveil information about applications, their configurations and environments, and potential changes between executions of the same applications.

\paragraph{Use Cases of Software Labels}
A fundamental application-specific insight is the identity and information about the software executing on HPC systems. 
Access to labeled data about software/application enables a range of use cases:
(a) Investigation of performance variability~\cite{costa2021systematically, feitelson2015workload, skinner2005understanding}.
(b) Application-specific system optimizations such as CPU frequency tuning~\cite{benkner2014automatic}, node sharing among compatible applications~\cite{breitbart2015case}, and node allocation based on application sensitivity to network~\cite{zhang2021using}.
(c) Detection of misuse or deviation from computing resource allocation purpose~\cite{ates2018taxonomist}.
(d) Estimation and prediction of the jobs' energy consumption~\cite{yamamoto2018classifying}.
(e) Application-specific statistics in system usage reports and guiding future system design decisions~\cite{ates2018taxonomist, yamamoto2018classifying}.

\paragraph{Problem Statement and Challenges}
HPC system operators generally do not have specific information about the executed applications beyond job names and executable names. 
Other information that users have shared voluntarily, such as application installation requests or user surveys regarding software usage, may be incomplete and relies on users reporting it accurately.
Typical indicators of application identity include static information such as Slurm job names and application binary file names~\cite{costa2021systematically}. 
One challenge is that these identifiers can be easily and arbitrarily changed by users~\cite{ates2018taxonomist}. 
Another challenge is that they do not reflect variations in application properties, such as compilation with various compilers, versions, or flags.
In the case of Python, the executable name will only give information about the name of the Python interpreter itself.

In contrast, application recognition based on dynamic information, such as resource use, can incur overhead associated with system monitoring or may fail in the presence of significant system noise or previously unseen application variations. 
HPC systems using services like LMOD for module environments can track the use of modules by jobs. 
Specific modules may be indicators of software use; however, they are unreliable because modules can be loaded by default, as dependencies of other loaded modules, or as a result of copy-pasting job scripts. 
Moreover, user-compiled software might not be tied to specific modules at all.

\paragraph{Limitations of Existing Solutions}
Existing data collection frameworks aim to overcome the problem of unknown software by collecting information at the process level of applications, such as XALT~\cite{agrawal2014user, mclay2015understanding} and LDMS with process tracking~\cite{agelastos2014lightweight, shoga2024evolving}. 
Collected information includes process metadata and environment information, such as the path to the executable, a cryptographic hash of the executable, loaded modules, loaded shared objects (libraries), and memory-mapped files.

Processes can execute different executables that still belong to the same software distribution. 
Even the \textit{same} application can appear under different versions due to compilation or code changes. 
Available data collection frameworks lack a fast and scalable way of collectively and comparatively labeling these application instances based on their similarities and the software they belong to.

\paragraph{Static Code Analysis}
Static code analysis approaches have been proposed for more reliable software identification and recognition in the context of HPC security~\cite{haridas2020code, Peisert2017-CACM-HPC} and HPC security action plans~\cite{nist2016action}.

Hash-based signatures of application executables have been proposed for software and library tracking~\cite{agrawal2014user, mclay2015understanding}.
Application signatures can also be used to collect application-specific resource usage data~\cite{yamamoto2018classifying}, and in secure identity frameworks as workload identifiers~\cite{spirespiffe}.

Compared to classical cryptographic hashing, similarity-preserving fuzzy hashing of application executables has also been investigated in the context of HPC application classification~\cite{jakobsche2024using}.

\paragraph{Fuzzy Hashing and SSDeep}
In cryptographic hashing, small changes in the input (e.g., small code changes to an executable application) lead to large changes in the resulting cryptographic hash; this is called the \textit{'avalanche effect'}~\cite{feistel1973cryptography}.
In contrast, a fuzzy hash is composed of multiple rolling hashes over parts of the input, preserving structural information and enabling a similarity-based comparison between fuzzy hashes. 
This means that two files that only differ slightly in content will also produce two fuzzy hashes that only differ slightly. 

While well-established in cybersecurity (as exemplified by the Microsoft 365 Defender research~\cite{garcia2021fuzz}), fuzzy hashing has only recently been adopted in HPC~\cite{jakobsche2024using}. 
Fuzzy hashing preserves privacy and integrity of the input files, and comparing fuzzy hashes is faster and more scalable than comparing files byte-by-byte.

\paragraph{Proposed Solution}
This work introduces the Software Identification and Recognition (SIREN) data collection framework, which collects process and executable file metadata, environment information, SSDeep fuzzy hashes of (a)~raw application binaries, (b)~printable strings of the binary, and (c)~the symbol table (comprising function and variable names).

SIREN collects these data using an LD\_PRELOAD-based hooking approach, injecting data collection code into processes.
Incorporating fuzzy hashes allows identification and recognition of application executables that belong to the same software, even if they are not exactly the same executable due to code changes, different versions, and/or compilers. 

\paragraph{Expected Impact}
Identifying and recognizing software and applications, as well as their metadata and execution environments, is crucial to ensuring efficient and secure use of HPC resources, as it allows a variety of investigations, such as current system use, application-specific performance variability, system optimization, misuse detection and/or deviation from allocation purpose, and energy consumption prediction. 
This information also helps reduce the HPC support response time for troubleshooting and guides future system design and procurement decisions based on software usage statistics.

\paragraph{Deployment and Analysis}
We deployed the SIREN framework for early testing on LUMI, collecting application data using a user-opt-in approach.
To highlight the analysis enabled by SIREN, we present in this work an analysis of anonymized user data, along with the framework architecture.
The data collected originates from 12 opt-in users, a total of 13,448 jobs, and 2,317,859 processes.
The analysis includes investigating typical information, such as loaded shared objects (libraries), and similarity-based identification and recognition of execution of initially \textit{unknown} and post-analysis \textit{known} software.

\paragraph{Main Contributions}
The main contributions of this work are:
\begin{itemize}
    \item Introduction of the SIREN data collection framework, which combines the collection of process-level metadata and environment information with the analysis of fuzzy hashes of application executables for the purpose of identifying and recognizing software usage on HPC systems.
    \item Analysis and findings from a first opt-in SIREN deployment on the LUMI supercomputer, highlighting SIREN's functionality to generate application-specific insights.
\end{itemize}

\paragraph{Paper Organization}
Section~\ref{section:background} provides background on the LUMI supercomputer, the LD\_PRELOAD-based data collection, and SSDeep fuzzy hashing.
Section~\ref{section:siren-overview} introduces the software architecture and components of the SIREN framework.
Section~\ref{section:analysis-findings} shows the analysis and findings from the first opt-in deployment of SIREN on the LUMI supercomputer.
The data collection frameworks related to this work are discussed in Section~\ref{section:related-work}, while Section~\ref{section:conclusion} concludes the work.

\section{Background} \label{section:background}
This section briefly introduces the LUMI supercomputer and describes the LD\_PRELOAD process hooking and fuzzy hashing with SSDeep.

\paragraph{The LUMI Supercomputer}
LUMI is an HPE Cray EX system and EuroHPC JU's flagship supercomputer. 
It is ranked \#8 on the 64th edition of the top500 and \#25 on the 24th edition of the Green500 lists. 
Its CPU partition has approximately 260,000 \texttt{AMD EPYC CPU} cores, while
its GPU partition consists of approximately 12,000 \texttt{AMD MI250X GPUs} with a sustained computing power of 380 Pflop/s (HPL $R_{max}$). 
LUMI has approximately 200 active users per month and uses the Slurm~\cite{yoo2003slurm} job scheduler.

\paragraph{Process Hooking with LD\_PRELOAD}
\texttt{LD\_PRELOAD} is an environment variable used by the Linux dynamic linker (\texttt{ld.so}) that allows a specified shared object to be injected into every newly launched process. 
By adding a custom library to the LD\_PRELOAD path, the linker loads it \textit{before} any other libraries, enabling both, adding new code to \textit{constructor} or \textit{destructor} and function interposition. 
The former uses \textit{constructor} and \textit{destructor} functions (annotated with \texttt{\_\_attribute\_\_((constructor))} and \texttt{\_\_attribute\_\_((destructor))} in C) that run before \texttt{main()} and upon process termination. 
The latter replaces references to the original library calls (e.g., \texttt{open()}) with equivalent custom routines that can log or alter parameters before forwarding them to the real functions. 

This \textit{LD\_PRELOAD}-based process hooking approach is appealing in the HPC settings because it operates without requiring modifications to user applications or system binaries, imposing minimal setup overhead. 
However, statically linked binaries (which do not invoke dynamic linking) cannot be intercepted.

\subsection{Fuzzy Hashing with SSDeep}
SSDeep uses a Context-Triggered Piecewise Hashing (CTPH)~\cite{kornblum2006identifying} to generate a \textit{fuzzy hash} for a given input. 
Unlike traditional \textit{cryptographic hashing}, the result of which changes completely if even a single bit changes, fuzzy hashing is designed to capture the overall structure of a file so that similar files produce similar \textit{fuzzy hashes}. 

SSDeep breaks a file into chunks based on its content (instead of fixed-size blocks); each chunk is independently hashed and then concatenated into a single fuzzy hash. 
The resulting fuzzy hashes can be compared using the Damerau–Levenshtein distance~\cite{damerau1964technique}, counting the necessary insertions, deletions, substitutions, or transpositions of two adjacent characters.
to make two strings the same; this is a variant of the Levenshtein distance~\cite{levenshtein1966binary} (counting only the insertions, deletions, or substitutions). 
SSDeep then converts these differences into a score between 0 and 100, where 0 indicates that the files are not similar at all and 100 that the files are effectively identical. 
Through this approach, SSDeep can determine how similar two fuzzy hashes are even if they are not identical.

By focusing on similarity rather than exact matching, SSDeep can identify files that are similar (e.g. slightly modified files). 
This also reduces the comparison time for large numbers of files at the byte level. 
Fuzzy hashes are traditionally used in malware detection; however, they have also recently been used to classify executable files in HPC systems~\cite{jakobsche2024using}. 

\section{SIREN Overview} \label{section:siren-overview}

SIREN is a process-level data collection framework for \textbf{s}oftware \textbf{i}dentification and \textbf{re}cognitio\textbf{n}. 
SIREN allows tracking and correlation of software use starting from \textit{libraries} (user and system) to application \textit{executables} submitted as \textit{jobs} by \textit{users} on an HPC system.

\begin{figure}[h]
    \centering
    \includegraphics[width=0.5\linewidth]{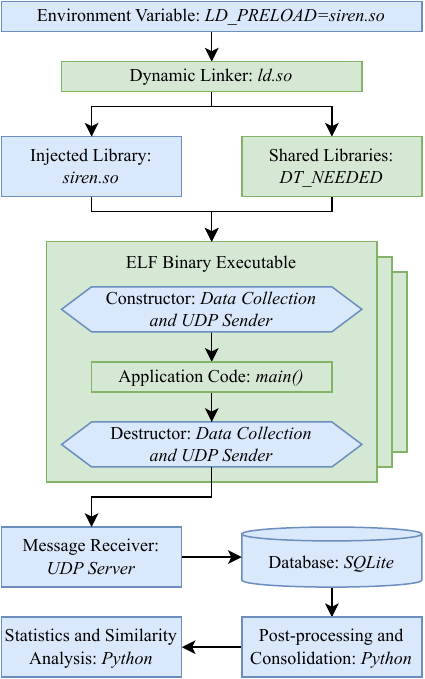}
    \caption{Process hooking mechanism through LD\_PRELOAD, injecting the SIREN data collection library into the constructor called at process initialization and the destructor called at process finalization. 
    Injected data collection code sends UDP messages to a message receiver, which saves data in an SQLite database. 
    Blue-shaded elements are part of the SIREN architecture.}
    \label{figure:siren-architecture}
\end{figure}

To observe and record information on HPC jobs and executables without modifications, we use the LD\_PRELOAD mechanism (see Section~\ref{section:background}) to load our custom-built shared object (\texttt{siren.so} written in C) into the target processes. 
This injected library executes code in the \textit{constructor} and \textit{destructor} functions, ensuring that data collection occurs at process start-up and before termination. 
All collected data is aggregated into formatted strings and sent via chunked UDP messages to a central database for logging and further analysis.
Figure~\ref{figure:siren-architecture} shows the architecture of the \texttt{siren.so} data collection library.
The \texttt{siren.so} data collection library is available online on Zenodo (\url{https://doi.org/10.5281/zenodo.15212885}) and is described in the accompanying artifact description form.

\paragraph{Module Deployment}
SIREN is deployed as a system module on LUMI written in Lua.
The module exports the LD\_PRELOAD path to the data collection library (\texttt{siren.so}). 
This deployment approach enables opt-in data collection, with users willingly participating in data collection by loading the SIREN module in their jobs.
We conducted a first opt-in deployment campaign on LUMI with 12 users between December 11, 2024 and March 7, 2025.

\subsection{Data Collection through \texttt{siren.so}}
SIREN's data collection is designed to be lightweight with a minimal set of dependencies.
The core functionality is handled through standard C libraries such as \texttt{<stdio.h>}, \texttt{<stdlib.h>}, \texttt{<string.h>}, and \texttt{<unistd.h>}.
However, for specialized functionality, the framework relies on a few external libraries:
the \texttt{libelf} library for processing and analyzing ELF binaries, 
the \texttt{xxHash} library~\cite{xxhash} for fast hash computations, and the \texttt{libfuzzy} library for the fuzzy hashing functionality of SSDeep.

Below is an overview of the primary data categories and their extraction method; 
if the executable is a Python interpreter, SIREN also collects a subset of this information for Python scripts.

\paragraph{Job and Process Identifiers} 
SIREN records the values of environment variables such as SLURM\_JOB\_ID, SLURM\_STEP\_ID, SLURM\_PROCID and HOSTNAME. 
SIREN collects the process ID (PID), the parent PID (PPID), the user ID (UID), and the group ID (GID) through system calls.
The path to the executable is obtained through \texttt{/proc/self/exe}, which is then hashed using a 128-bit hash (XXH3\_128bits from the xxHash library). 
The hash produced by xxHash is neither a cryptographic nor a fuzzy hash and is optimized only for performance.
It is used to separate PID collisions in the database (e.g., bash replacing itself with \texttt{exec()} or \texttt{execv()}) and is not further analyzed.
SIREN also collects executable file metadata such as inode number, file size, file permissions, file owner UID, file owner GID, as well as access, modification, and change timestamps.

\paragraph{Modules, Libraries, Compilers, and Mapped Memory} 
SIREN collects the list of modules loaded from the environment variable \texttt{LOADEDMODULES}.
The loaded shared objects (libraries) of the executables are extracted through \texttt{dl\_iterate\_phdr} which is part of the standard C library.
During compilation, most compilers leave an identification string in the \texttt{.comment} section of the executable; 
we use the \texttt{libelf} library to parse this section and extract any stored compiler identification strings.
SIREN also extracts the memory-mapped regions of the process by parsing \texttt{/proc/self/maps}; 
for Python interpreters, this information is later post-processed to extract imported packages.

The list of modules, libraries, compiler information, and mapped memory information is additionally fuzzy hashed with SSDeep to provide a means of analysis and similarity, even in the case of partially missing information due to UDP message loss.

\paragraph{Executables, Python Scripts, and Fuzzy Hashes} 
SIREN uses SSDeep to compute fuzzy hashes, which help identify similarities between executables or scripts and include hashing of:
The content of the raw executable file, the printable strings found in the file (similar to the output of the \texttt{strings} command), the global scope ELF symbols (similar to the output of the \texttt{nm} command) and the Python input script (if present).
The global scope of ELF symbols refers to externally visible functions and variables defined without the \textit{static} keyword, meaning that they form part of the public interface of an application. 
They are more suitable for application identification because they consistently expose the essential identifiers and functionalities of an application.

\paragraph{Selective Data Collection} \label{section:selective-data-collection}
We collect data only for the process with \texttt{SLURM\_PROCID=0}, to avoid collecting duplicate data in the case of multiple MPI ranks. 
Processes are divided according to where their executables originate from, into the categories \textit{system}, \textit{user}, and additionally \textit{Python}.
\textit{System} is the process category that executes executables from system directories (\texttt{/etc/, /dev/, /usr/, /bin/, /boot/, /lib/, /opt/, /sbin/, /sys/, /proc/, /var/}).
\textit{User} is the process category that executes executables \textit{not} from system directories.
\textit{Python} is the process category that executes a Python interpreter from a system directory; 
if the Python interpreter is installed in a user directory, it will count as a \textit{user} executable.
We selectively skip collection (or only collect certain information) for executables per each category to reduce overhead and redundant data collection, as shown in Table~\ref{table:data-collection}.
For example, it is unnecessary to repeatedly hash an executable like \texttt{bash} from the \texttt{/usr/bin/} system directory.

\begin{table}[h]
    \centering
    \begin{tabular}{l|cccc}
        \toprule
        \textbf{Collected}       & \textbf{System}        & \textbf{User}          & \textbf{Python}        & \textbf{Python}  \\
        \textbf{Information}     & \textbf{Executable}    & \textbf{Executable}    & \textbf{Interpreter}   & \textbf{Script} \\
        \midrule
        File Metadata   & \checkmark           & \checkmark           & \checkmark           & \checkmark \\
        Libraries       & \checkmark           & \checkmark           & \checkmark           & \xmark\ \\
        Modules         & \xmark\            & \checkmark           & \xmark\            & \xmark\ \\
        Compilers       & \xmark\            & \checkmark           & \xmark\            & \xmark\ \\
        Memory Map      & \xmark\            & \checkmark           & \checkmark           & \xmark\ \\
        File\_H         & \xmark\            & \checkmark           & \xmark\            & \checkmark \\
        Strings\_H      & \xmark\            & \checkmark           & \xmark\            & \xmark\ \\
        Symbols\_H      & \xmark\            & \checkmark           & \xmark\            & \xmark\ \\
        \bottomrule
    \end{tabular}
    \caption{Data collection for different scopes.}
    \label{table:data-collection}
\end{table}

We collect the memory-mapped files of Python interpreters to extract the imported Python packages in post-processing.
For Python scripts, we only collect the file metadata and a fuzz hash of the script itself.
Information such as libraries and compilers cannot be extracted in this case, because Python scripts are not compiled binaries.

\paragraph{Data Transmission} 
One main goal of SIREN is to \textit{gracefully fail} in the presence of data collection errors so that user processes remain unaffected by the data collection mechanism.
A primary design concern was to minimize interference with application processes, especially when hooking onto a potentially large number of processes via LD\_PRELOAD.
For these reasons, we decided for a UDP-based approach over TCP or file-based methods (such as creating individual files for every hooked process).
The \textit{'fire and forget'}~\cite{roehle1997channeling} messaging approach of UDP avoids the overhead of connection management and the resource burden of file-based logging, minimizing additional complexity and potential failure points due to excessive open file handles or socket exhaustion.
Even when processes stop unexpectedly, in the event of transient network errors, or errors in the data collection itself, SIREN can simply fail gracefully.

\paragraph{UDP Message Sender} \label{section:udp-sender}
The SIREN data collection library (\texttt{siren.so}) includes a UDP message sender. 
Multiple UDP messages are sent for each category (e.g. modules, shared libraries, compilers, etc.). 
Since UDP messages are limited in size, the UDP sender additionally chunks longer messages (e.g., lists of modules and shared libraries) and sends these chunks separately. 

The UDP messages have an additional header to distinguish individual processes in the database. 
The header fields are as follows: 
JOBID, STEPID, PID, HASH (a hash of the path to the executable), HOST (hostname of the node), TIME (the \texttt{UNIX} timestamp of data collection), LAYER (SELF or SCRIPT to distinguish Python interpreters from Python scripts), TYPE (the type of information, e.g. MODULES, OBJECTS, COMPILERS), and CONTENT (the content of the message, e.g. the list of loaded modules). 

\paragraph{UDP Message Receiver and Database} 
The receiver server is written in Go.
Whenever incoming messages are received, they are read from
the buffered channel of the receiver server and then inserted into the SQLite database.
Since SIREN collects individual information as UDP messages (e.g. shared libraries), parts of this information can be lost due to UDP message loss.
For our first deployment, approximately 0.02\% of the jobs have missing fields that can be attributed to the loss of UDP messages.

The SQLite database stores UDP messages based on columns (UDP header information): JOBID, STEPID, PID, HASH, HOST, TIME, LAYER, TYPE, and CONTENT.

\paragraph{Differentiating duplicate PIDs for different executables} 
Under certain circumstances, different executables can be executed under the same JOBID, STEPID, HOST, TIME, and PID. 
This problematic situation arises when a process calls an \texttt{exec()}-family function, which replaces the current process image with a new one while retaining the same PID. 
Since \texttt{UNIX} timestamps have a one-second granularity, it is possible to see the same PID associated with different executables with the same timestamp. 
For this particular reason, SIREN additionally collects the hash of the path to the executable as an identifier. 
This situation should not be confused with PID collisions (reusing PIDs of old processes), which, in reality, is highly unlikely to occur within the same timestamp. However, this is also covered by hashing the path of the executable.

\paragraph{Post-processing and Analysis} 
Post-processing of UDP messages from the database includes the merging of multiple UDP message chunks into single data records per process.
Information about Python scripts is merged into their parent (Python interpreter) rows. 
The result is a single database entry for each process, including information about the input scripts in the case of Python.
After postprocessing UDP messages, we analyze the consolidated rows of process information through Python scripts.

\paragraph{Requirements and Limitations} 
Apart from setting the LD\_PRELOAD environment variable and setting up the UDP message receiver server and database, there are no other particular requirements for using SIREN.
However, SIREN does not currently collect information about statically linked executables (since these do not invoke dynamic linking), nor information about processes inside containers.
The LD\_PRELOAD environment variable is propagated within the container. 
However, the directory containing the shared library (\texttt{siren.so}) is not currently automatically mounted within the container.

\section{Analysis and Findings} \label{section:analysis-findings}
This section presents the analysis of the data collected on LUMI using the opt-in deployment campaign.

\subsection{Users, Jobs, and Processes}
A cohort of 12 users participated in our deployment campaign, executing a total of 13,448 jobs and 2,317,859 processes between December 11, 2024 and March 7, 2025, described in Table~\ref{table:overview}.
User names are anonymized by random assignment of \textit{user\_<int>} to UIDs. 
Separation of processes into those originating from system or user directories or Python is achieved according to the criteria listed in Section~\ref{section:selective-data-collection}.

The data in Table~\ref{table:overview} reveal different usage patterns between participants, with certain users executing system executables exclusively and others combining them with user executables or Python scripts. 
User \textit{user\_1} submitted over 11,000 jobs with more than 1.7 million processes using only executables from system directories (such as \texttt{/usr/bin/mkdir} and \texttt{/usr/bin/rm}).
Most other users employ a mix of executables from the system and the user directory, which is expected since executables such as \texttt{/usr/bin/lua} (for module loading) and \texttt{/usr/bin/srun} are typically present in every job.
More than 23,000 Python processes were launched by \textit{user\_4}.
In contrast to other Python users, \textit{user\_4} employs a mix of Python and user directory executables.
The data from \textit{user\_6} present an interesting case, since they do not use any executables from the system directories, indicating that there is no use for \texttt{/usr/bin/srun} or \texttt{/usr/bin/lua}.

\begin{table}[h]
    \centering
    \begin{tabular}{l|rrrr}
        \toprule
        \textbf{User}         & \textbf{Job}    & \textbf{System Dir.}   & \textbf{User Dir.} & \textbf{Python}    \\
        \textbf{(anonymized)} & \textbf{count}  & \textbf{Processes}     & \textbf{Processes} & \textbf{Processes} \\
        \midrule
        user\_1      & 11,782   & 1,731,077     & --        & -- \\
        user\_2      & 930      & 48,095        & 5,259     & -- \\
        user\_11     & 230      & 3,980         & 138       & -- \\
        user\_8      & 216      & 3,039         & 2,103     & -- \\
        user\_4      & 205      & 528,205       & 642       & 23,286 \\
        user\_5      & 47       & 94            & --        & 29 \\
        user\_10     & 28       & 3,336         & 889       & -- \\
        user\_9      & 4        & 8             & 4         & -- \\
        user\_3      & 2        & 6             & 4         & -- \\
        user\_6      & 2        & --            & 2         & -- \\
        user\_7      & 1        & 17            & 1         & -- \\
        user\_12     & 1        & 2             & --        & 1 \\
        \midrule
        Total        & 13,448   & 2,317,859     & 9,042     & 23,316 \\
        \bottomrule
    \end{tabular}
    \caption{Data about users' jobs and processes. Rows sorted in descending order by job count, system directory process count, user directory process count, and Python process count.}
    \label{table:overview}
\end{table}

\subsection{System Directory Processes}
Executables originating from system directories are usually installed by system administrators. 
The data in Table~\ref{table:system_executables} denote the top 10 executables most frequently used in the system directories. 
The total number of executables in the system directories is 112. 
\texttt{OBJECTS\_H} is an SSDeep fuzzy hash of the list of loaded shared objects (libraries).

\begin{table}[h]
    \centering
    \begin{tabular}{lrrrr}
        \toprule
        \textbf{Executable}      & \textbf{Unique} & \textbf{Job}   & \textbf{Process} & \textbf{Unique} \\
        \textbf{Path \& Name}    & \textbf{Users}  & \textbf{Count} & \textbf{Count}   & \textbf{OBJECTS\_H} \\
        \midrule
        /usr/bin/srun   & 10           & 1,642   & 4,564     & 3 \\
        /usr/bin/bash   & 8            & 13,105  & 161,418   & 3 \\
        /usr/bin/lua5.3 & 8            & 882     & 18,448    & 2 \\
        /usr/bin/rm     & 6            & 12,182  & 544,025   & 1 \\
        /usr/bin/cat    & 6            & 9,774   & 29,003    & 1 \\
        /usr/bin/uname  & 5            & 1,182   & 28,053    & 1 \\
        /usr/bin/ls     & 5            & 1,130   & 9,057     & 1 \\
        /usr/bin/mkdir  & 4            & 8,863   & 547,089   & 1 \\
        /usr/bin/grep   & 4            & 1,115   & 9,268     & 1 \\
        /usr/bin/cp     & 4            & 1,019   & 11,655    & 1 \\
        \bottomrule
    \end{tabular}
    \caption{Top 10 most used executables from system directories. Rows sorted in descending order by unique users, job count, process count, and unique OBJECTS\_H count.}
    \label{table:system_executables}
\end{table}

The executable \texttt{/usr/bin/srun} is used by most, but not all, of the 12 participants, indicating the presence or absence of \texttt{srun} in their job scripts, respectively.
Similarly, \texttt{/usr/bin/lua} (for module loading) is not used by all users.
The high number of processes executing \texttt{/usr/bin/mkdir} and \texttt{/usr/bin/rm} can be attributed to \textit{user\_1}, who seems to rely heavily on executables from system directories, as illustrated in Table~\ref{table:overview}.

The presence of multiple unique instances of OBJECTS\_H in the data shown in Table~\ref{table:system_executables} indicates that the same executables are executed as \textit{variants} that load distinct sets of shared libraries (which can influence their execution behavior).

This behavior is typical for HPE (Cray) software such as the Cray Programming Environment, which includes multiple wrappers for compilers. 
The behavior of these wrappers is controlled by loading specific modules together with the modules for the compilers. This affects which libraries are dynamically linked against the executable at compilation time, also requiring the respective modules to be loaded at execution time.

\paragraph{Deviating Shared Libraries}
Table~\ref{table:bash_objects} shows the different set of shared libraries for different variants of \texttt{/usr/bin/bash}. 
The 3 different unique \textit{OBJECT\_H} in Table~\ref{table:system_executables} are caused by the different versions of \texttt{libtinfo} and \texttt{libm} loaded by \texttt{/usr/bin/bash} in different user environments.

The \texttt{libtinfo} library belongs to ncurses (a library used to write text-based user interfaces) and enables querying and managing terminal capabilities. 
Developers can link directly with \texttt{libtinfo} if they need only the terminal capability functions without the overhead of the full ncurses user interface functions.
The \texttt{libm} library is the standard mathematical library on Linux. 
It provides a wide range of mathematical functions that are essential for scientific computations, engineering applications, and advanced math operations.

\begin{table}[h]
    \centering
    \begin{tabular}{lrrr}
        \toprule
        \textbf{Executable} & \textbf{Processes} & \textbf{libtinfo Path} & \textbf{libm Path} \\
        \midrule
        /usr/bin/bash   & 160,904     & /lib64/libtinfo.so.6          & -- \\
        /usr/bin/bash   & 460         & .../spack/.../libtinfo.so.6   & -- \\
        /usr/bin/bash   & 54          & .../SW/.../libtinfo.so.6      & /lib64/libm.so.6 \\
        \midrule
        Total           & 161,418     & & \\
        \cmidrule(lr){1-2}
    \end{tabular}
    \caption{Distinct Sets of Shared Objects (Libraries). Rows sorted in descending order by process count.}
    \label{table:bash_objects}
\end{table}

Bash itself typically does not need \texttt{libm}, though it can link to it if it is built with certain features (like advanced arithmetic expansions or optional loadable built-ins). 
A custom library or an environment module might bring in \texttt{libm} as a dependency, causing bash to load it even if it is not strictly required by its default configuration.
However, in most HPC scenarios, \texttt{libm} is more likely for the use of the \texttt{bc} bash calculator in a script, a command-line utility in Unix-like operating systems that stands for "basic calculator."
\texttt{bc} is used within bash scripts and interactive shell sessions to perform arithmetic calculations, especially for non-integer (floating point) arithmetic.

The behavior of system-installed executables can be altered by loading different libraries due to the user environment settings.
Analyzing and detecting deviations from commonly loaded libraries helps the user support team troubleshoot situations where standard tools are reported (for example, in user support tickets) to behave unexpectedly.

\subsection{User Directory Processes}
The nature of executables originating from user directories is generally \textit{unknown}.
However, system operators can often deduce to which software an executable belongs based on file or path names by using regular expressions to match with known software names~\cite{turner2022software}.

In the case of nondescript file or path names, or in the case of Python, the identity of executed software typically remains unknown.
The names of executables can also collide, for example, by repeatedly reusing compilation commands that generate executables called \texttt{a.out} that may belong to entirely different applications.

\paragraph{Derived Labels for User Applications}
Table~\ref{table:derived_labels} shows software labels for user applications derived from executable file or path names based on regular expression matching.

\begin{table}[h]
    \centering
    \begin{tabular}{lrrrr}
        \toprule
        \textbf{Software}              & \textbf{Unique} & \textbf{Job}   & \textbf{Process} & \textbf{Unique} \\
        \textbf{Label}                 & \textbf{Users}  & \textbf{Count} & \textbf{Count}   & \textbf{FILE\_H} \\
        \midrule
        LAMMPS                  & 2            & 226    & 226       & 5 \\
        GROMACS                 & 2            & 215    & 2,104     & 1 \\
        miniconda               & 1            & 673    & 5,018     & 5 \\
        janko                   & 1            & 138    & 138       & 2 \\
        icon                    & 1            & 64     & 625       & 175 \\
        amber                   & 1            & 27     & 889       & 2 \\
        gzip                    & 1            & 18     & 19        & 1 \\
        \texttt{UNKNOWN}        & 1            & 3      & 17        & 7 \\
        alexandria              & 1            & 2      & 4         & 1 \\
        RadRad                  & 1            & 2      & 2         & 2 \\
        \bottomrule
    \end{tabular}
    \caption{Derived Labels for User Applications. Rows sorted in descending order by unique users, job count, process count, and unique FILE\_H count, respectively.}
    \label{table:derived_labels}
\end{table}

All executable instances could be matched to a specific known software application or program, except one that had a nondescript path and file name, which we labeled \texttt{UNKNOWN}.
FILE\_H is the SSDeep fuzzy hash of the raw executable binary, where a perfect match means that it is effectively the same executable.
Multiple instances of (exactly) the same executable can exist in different paths, as long as compilation processes were employed that lead to the same compiled binary.
Table~\ref{table:derived_labels} shows the number of \textit{unique} hashes (last column), indicating the number of (even slightly) different executables per software package.

This information can be reported as usage statistics about specific software, as well as to investigate correlations of research domains (based on software executed) to the number of users, jobs, and processes.
Compared to the other instances, \texttt{LAMMPS} (Large-scale Atomic/Molecular Massively Parallel Simulator) and \texttt{GROMACS} (originally GROningen MAchine for Chemical Simulations, however, currently not an abbreviation) are used by multiple users.
This gives an indication of the diversity of different versions of software.
For example, \texttt{GROMACS} exists as a single unique instance executed by multiple users, while \texttt{icon} is found under multiple distinct executables used by a single user.
An interesting example is \texttt{gzip}, which appears to be a user-installed version of a compression utility.
The \texttt{UNKNOWN} case is further analyzed in Section~\ref{section:identyfing-software}.

\paragraph{Compiler Information of Applications in User Directories} \label{section:compilers}
An overview of the different compilers used by user applications is shown in Table~\ref{table:compilers}.
Compilers typically leave an identifier in the \texttt{.comment} section of ELF files; 
if the application executable is built from dependencies with different parts compiled by different compiler versions, this may result in a list of (the multiple involved) compilers.

\begin{table}[h]
    \centering
    \begin{tabular}{lrrrr}
        \toprule
        \textbf{Compiler Name [Provenance]} & \textbf{Unique} & \textbf{Job}  & \textbf{Process} & \textbf{Unique} \\
                 & \textbf{Users}  & \textbf{Count}     & \textbf{Count}          & \textbf{FILE\_H} \\
        \midrule
        LLD [AMD]                                   & 4 & 244 & 2,145 & 4 \\
        GCC [SUSE]                                  & 4 & 223 & 788   & 134 \\
        GCC [SUSE], clang [Cray]                      & 2 & 40  & 46    & 34 \\
        GCC [Red Hat], GCC [conda]                    & 1 & 673 & 4,983 & 4 \\
        GCC [SUSE], GCC [HPE]                         & 1 & 138 & 138   & 2 \\
        GCC [Red Hat], rustc                          & 1 & 35  & 35    & 1 \\
        GCC [SUSE], clang [AMD]                       & 1 & 27  & 889   & 2 \\
        GCC [SUSE], clang [Cray], clang [AMD]          & 1 & 18  & 18    & 13 \\
        \bottomrule
    \end{tabular}
    \caption{Compiler Information of Applications in User Directories. 
    Rows sorted in descending order by unique users, job count, process count, and unique FILE\_H count.}
    \label{table:compilers}
\end{table}

This information offers insights into the compilation of user applications.
The presence of multiple compiler references within a single executable indicates a combination of compiler usage or dependencies, each potentially built using a distinct toolchain.
This revelation can expose compatibility issues or explain performance variations when the same application is built repeatedly with varying configurations.
Furthermore, it discloses the emergence of novel toolchains (e.g., \texttt{Rust} or \texttt{conda-based GCC}) and shows the prevalence and diversity of these builds.

\paragraph{Scientific Shared Objects in User Applications} \label{section:libraries}
Application executables are usually built with a number of shared libraries.
The loaded shared objects (libraries) can be extracted and analyzed to provide insight into the application nature.
The presence of particular libraries can indicate the scientific purpose of an application (e.g. \texttt{climatedt} for climate/weather related codes).

\begin{figure}[h]
    \centering
    \includegraphics[width=\linewidth]{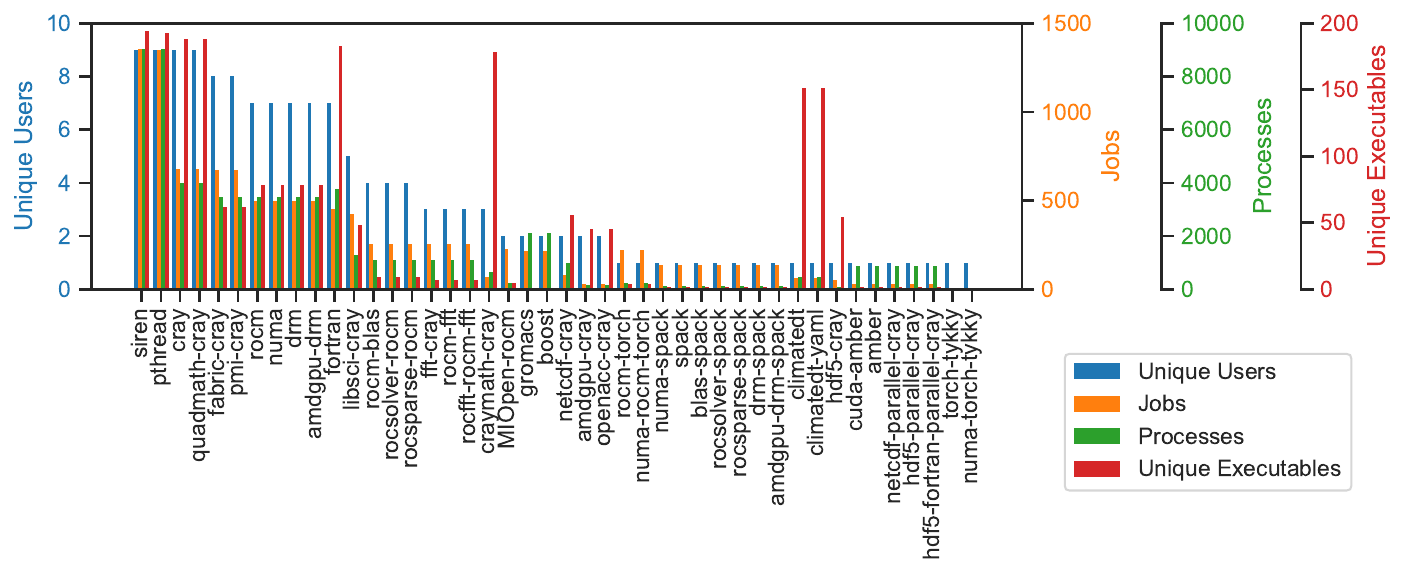}
    \caption{Derived and filtered shared objects (libraries) extracted from executables in user directories. 
    The four y axes denote the count of unique users, jobs, processes, and unique executables.
    As certain users do not launch executables from their user directory, the largest unique user count is less than the 12 participants.}
    \label{figure:shared-objects}
\end{figure}

Figure~\ref{figure:shared-objects} shows for each derived and filtered libraries found in user applications, the count of unique users, jobs, processes, and executables.
\textit{Derived and filtered} in our context means that we only extracted specific combinations of substrings of libraries, since the list of all shared libraries can be extremely long with many (potentially uninformative) libraries that provide basic functionality without giving insight into the application nature.
The list of combinatorial substrings extracted in our case is: \texttt{libsci, pthread, pmi, netcdf, hdf5, fortran, parallel, python, fabric, numa, boost, openacc, amdgpu, cuda, drm, rocsolver, rocsparse, rocfft, MIOpen, rocm, gromacs, blas, fft, torch, quadmath, craymath, cray, tykky, climatedt, amber, spack, yaml, java, siren.}

These results reveal that user applications are based on a mix of specialized libraries tied to different scientific domains and GPU architectures. 
For example, multiple ROCm-related libraries (\texttt{rocsolver}, \texttt{rocsparse}, \texttt{rocfft}, and \texttt{MIOpen}) point to AMD GPU usage, while \texttt{HDF5}, \texttt{NetCDF}, and \texttt{Fortran} libraries often appear in traditional climate and simulation codes.
The presence of libraries like \texttt{climatedt} or \texttt{gromacs} shows how this information can identify the application domain, e.g., weather simulation or molecular dynamics.

The significant disparity between the number of unique executables and the relatively low number of jobs and processes associated with \texttt{climatedt} and \texttt{climatedt-yaml} as depicted in Figure~\ref{figure:shared-objects} suggests that these libraries are used in a wide range of executables, while they are not integrated into numerous jobs or processes compared to other libraries.
Such scenarios can manifest, for instance, when multiple executables that rely on the same libraries are collectively part of a single job, resulting in a disproportionately elevated executable count relative to the job count.

\paragraph{Identifying Unknown Applications} \label{section:identyfing-software}
Table~\ref{table:ssdeep} shows the approach to identify an
unknown application executable with nondescript file and path name (denoted as \texttt{UNKOWN} introduced in Table~\ref{table:derived_labels}).
The columns show the similarity of various fuzzy hashes to the baseline fuzzy hash--that of the \texttt{UNKNOWN} instance. 
\textit{Avg. Sim.} denotes the average similarity across all columns,
\textit{MO\_H} the modules hash, 
\textit{CO\_H} the compilers hash, 
\textit{OB\_H} the shared objects (libraries) hash, 
\textit{FI\_H} the raw file hash, 
\textit{ST\_H} the printable strings hash, and
\textit{SY\_H} the function names hash.

Given \textit{UNKNOWN} as a baseline, we aim to identify the most similar \textit{known} case and compare their SSDeep fuzzy hashes in terms of similarity with respect to various characteristics of static files. 
These characteristics include loaded modules, compilers, shared libraries, raw binary, printable strings, and function names.
Using this approach, we discovered that the \textit{UNKNOWN} application exhibits a high degree of similarity to \textit{known} instances of \texttt{icon} (as introduced in Table~\ref{table:derived_labels}). 
Notably, one instance of \texttt{icon} exhibited a perfect 100\% similarity in all columns, while multiple other instances exhibited a decreasing similarity score.

\begin{table}[h]
    \centering
    \small
    \setlength{\tabcolsep}{4pt}
    \caption{Similarity Search Result for <unknown> case. Rows sorted in descending order by Average Similarity.}
    \begin{tabular}{lrrrrrrr}
        \toprule
        \textbf{Label} & \textbf{Avg. Sim.} & \textbf{MO\_H} & \textbf{CO\_H} & \textbf{OB\_H} & \textbf{FI\_H} & \textbf{ST\_H} & \textbf{SY\_H} \\
        \midrule
        \textbf{icon} & \textbf{100.0} & \textbf{100} & \textbf{100} & \textbf{100} & \textbf{100} & \textbf{100} & \textbf{100} \\
        icon & 94.8  & 96  & 100 & 100 & 83  & 90  & 100 \\
        icon & 83.7  & 94  & 100 & 57  & 68  & 83  & 100 \\
        icon & 81.3  & 82  & 100 & 57  & 69  & 80  & 100 \\
        icon & 77.8  & 100 & 100 & 100 & 0   & 85  & 82  \\
        icon & 77.8  & 100 & 100 & 100 & 0   & 85  & 82  \\
        icon & 74.8  & 96  & 100 & 100 & 0   & 71  & 82  \\
        icon & 74.8  & 96  & 100 & 100 & 0   & 71  & 82  \\
        icon & 67.0  & 94  & 100 & 57  & 0   & 69  & 82  \\
        icon & 67.0  & 94  & 100 & 57  & 0   & 69  & 82  \\
        \bottomrule
    \end{tabular}
    \label{table:ssdeep}
\end{table}

This reveals that the initially \textit{UNKNOWN} instance, which was executed with nondescript file and path names, is matched to \textit{known} instances.
In addition, it shows that specifically \texttt{icon} exists as multiple executable variants, potentially caused by different compilation settings, software versions, or manual code modifications.
Large software distributions can also include multiple distinct executables that are responsible for different functionalities or steps in the software workflow.
However, these executables belong to and are executed in the context of the same (overarching) software package.

\paragraph{Verifying Functionality of Scientific Software}
It is important to note that our labels (for \textit{known} software) were derived so far from file and path names.
This means that we can recognize an \textit{UNKNOWN} instance, as, for example, \texttt{icon}, only based on its similarity to other instances; 
this, however, does not yet establish what the software is \textit{actually} about.

Further analysis of the shared libraries can confirm this insight.
In the case of \texttt{icon} the derived and filtered shared objects are: \texttt{cray, craymath-cray, fortran, pthread, quadmath-cray, siren, climatedt, climatedt-yaml, fabric-cray, pmi-cray, hdf5-cray, netcdf-cray, libsci-cray, amdgpu-drm, drm, numa, rocm, amdgpu-cray, openacc-cray}.
From this list, we can derive the functionality of the software, with libraries like \texttt{HDF5}, \texttt{NetCDF}, \texttt{Fortran}, and especially \texttt{climatedt} (Destination Earth Digital Twin for Climate Change Adaptation~\cite{climateDT}) indicating that the functionality of software and the scientific domain are related to climate and weather simulations.

\subsection{Python Processes}
Python is a special case for identifying and recognizing executables on HPC systems, and hence, for our data collection library.
SIREN collects data at the process-level; 
for Python jobs, this targets the process executing the Python interpreter itself.
However, multiple users can potentially use the same interpreter simultaneously, which does not provide insight into what they actually execute.
We overcome this challenge by extracting the imported Python packages from the memory-mapped files of the Python interpreter.

\begin{table}[h]
    \centering
    \begin{tabular}{lrrrr}
        \toprule
        \textbf{Python} & \textbf{Unique}       & \textbf{Job} & \textbf{Process} & \textbf{Unique} \\
        \textbf{Interpreter}            & \textbf{Users}        & \textbf{Count}     & \textbf{Count}          & \textbf{SCRIPT\_H} \\
        \midrule
        python3.10  & 2            & 30   & 30        & 27 \\
        python3.6   & 1            & 28   & 14,884    & 6 \\
        python3.11  & 1            & 8    & 8,402     & 5 \\
        \bottomrule
    \end{tabular}
    \caption{Python Interpreters. Rows sorted in descending order by unique users, job count, process count, and unique SCRIPT\_H count.}
    \label{table:interpreters}
\end{table}

\paragraph{Python Interpreters}
Table~\ref{table:interpreters} shows the most executed Python interpreters by user, job, process, and unique script hash.
\textit{SCRIPT\_H}--the unique fuzzy hash of the Python input script--reveals the number of distinct Python scripts executed by each interpreter.
Our analysis reveals that all participants used different versions of \texttt{Python 3} with most processes executing \texttt{Python 3.6}, and greatest diversity with respect to different Python input scripts for \texttt{Python 3.10}.

\begin{figure}[h]
    \centering
    \includegraphics[width=\linewidth]{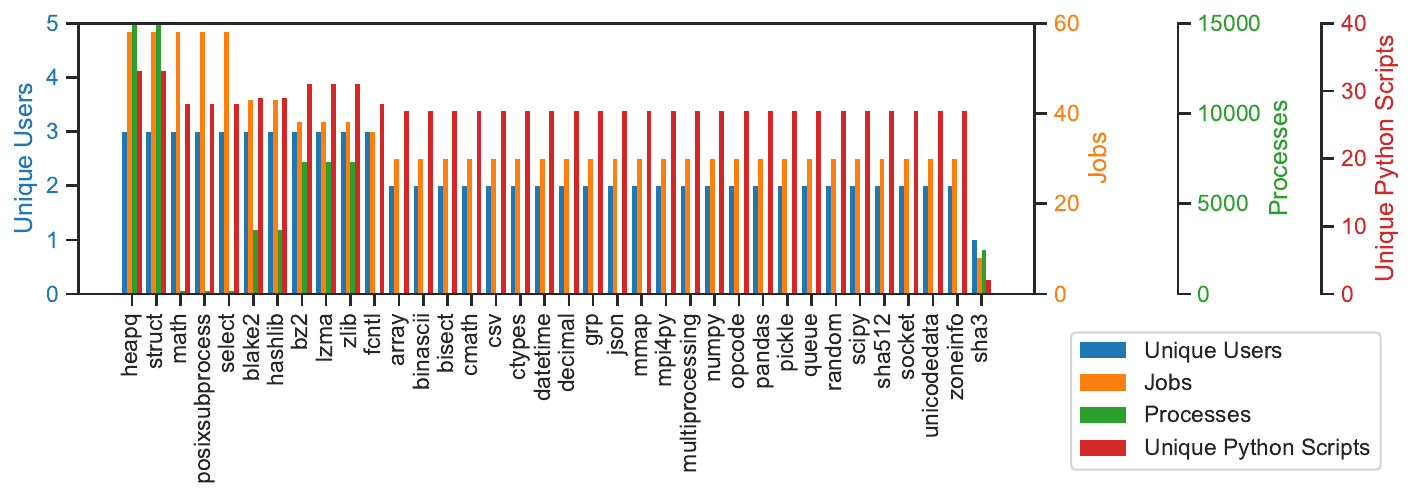}
    \caption{Imported packages extracted from the memory-mapped files of Python interpreters. The four y axes denote the count of unique users, jobs, processes, and unique Python scripts.}
    \label{figure:python-imports}
\end{figure}

\paragraph{Python Imported Packages}
Figure~\ref{figure:python-imports} shows for each imported Python package the unique users, jobs, processes, and Python scripts.
Based on the data collected on imported Python packages, we derive several insights regarding the use of Python.
Packages such as \texttt{heapq} and \texttt{struct} are imported by the three Python users, indicating that these packages are basic components in almost every Python execution for our participants.
Other packages like \texttt{mpi4py}, \texttt{numpy}, \texttt{pandas}, and \texttt{scipy} appear less frequently, imported by only a subset of Python users, suggesting their use in more specialized cases.
The count of unique Python scripts provides insight into Python script diversity, indicating whether the same or different Python script(s) are always used.

Investigating imported Python packages is crucial to detect potential Common Vulnerabilities and Exposures (CVEs)~\cite{pyupio_safetydb}, particularly in light of the increasing adoption of large language models (LLMs) for code generation.
The recent trend towards using LLMs to generate code has brought forth a novel security vulnerability known as 'slopsquatting'~\cite{claburn2025llms}.
Slopsquatting involves LLMs hallucinating nonexistent Python package names and malicious actors fabricating packages under these hallucinated names and then uploading them to package registries or indices such as PyPI or npm. 
Consequently, unsuspecting users install and execute these packages, potentially exposing themselves to malicious code.
This emphasizes the imperative to thoroughly investigate the use of software and imported Python packages to mitigate such vulnerabilities.

\subsection{Dependency Analysis}
Software packages are dependent on various compilers and scientific libraries.
In the analysis below, we examine software dependencies with regard to compilers and libraries.

\begin{figure}[h]
    \centering
    \includegraphics[width=0.4\linewidth]{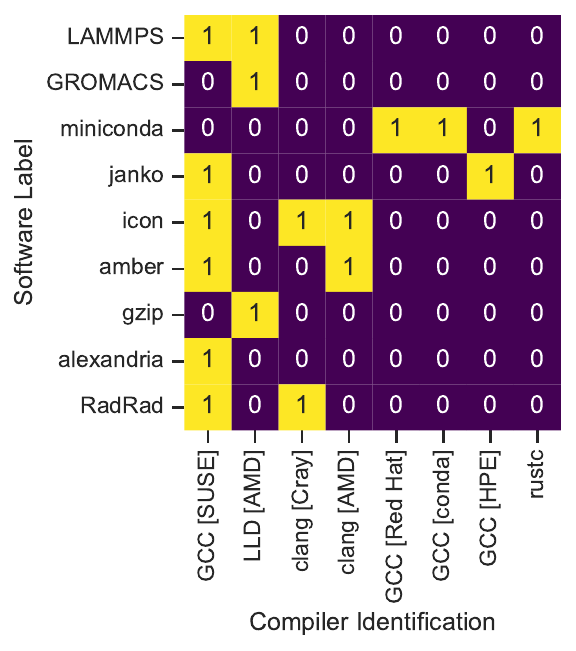}
    \caption{Compiler identification strings by software label (1=usage, 0=no usage).}
    \label{figure:labels-and-compilers}
\end{figure}

\paragraph{Compilers}
As discussed in Section~\ref{section:compilers}, the build process of an executable may involve multiple compilers.
Figure~\ref{figure:labels-and-compilers} provides an overview of the compiler identification strings found in each group of executables (based on software labels).
This analysis provides insight into the usage and dependencies of specific compilers, enabling informed decisions about updating or maintaining specific compilers and toolchains on the HPC system.

\begin{figure}[h]
    \centering
    \includegraphics[width=\linewidth]{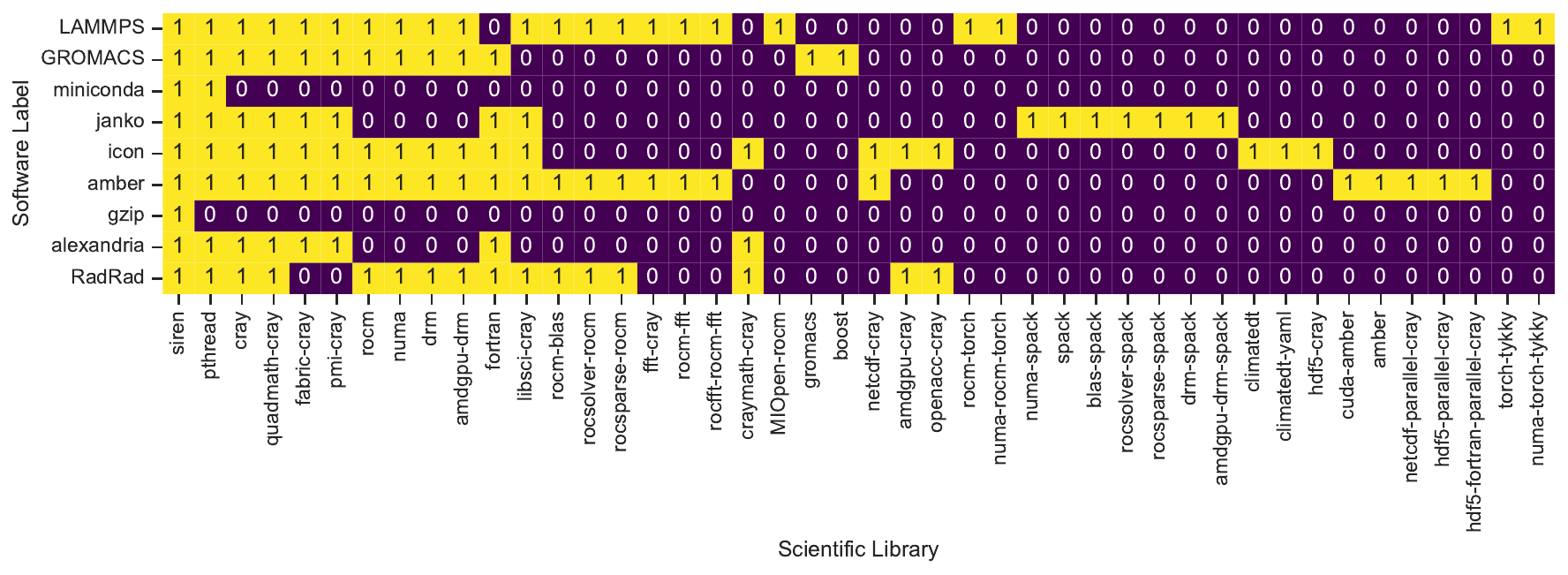}
    \caption{Loaded shared object (library) usage by software label (1=usage, 0= no usage).}
    \label{figure:labels-and-libraries}
\end{figure}

\paragraph{Libraries}
Figure~\ref{figure:labels-and-libraries} provides an overview of the libraries used by various software applications. 
This representation is a filtered and derived subset of libraries, as outlined in Section~\ref{section:libraries}.
The analysis of this visualization offers valuable insight into the specific library usage patterns of software applications. 
For example, it highlights which applications rely on specific versions of the HDF5 library (\texttt{hdf5-cray}, \texttt{hdf5-fortran-parallel-cray}, or \texttt{hdf5-parallel-cray}).
The \texttt{siren(.so)} library is loaded by all executables because it is the shared library injected through LD\_PRELOAD that the SIREN framework uses for data collection.
Other essential components include the POSIX \texttt{pthreads} library;
it is leveraged either directly or indirectly (e.g., through OpenMP~\cite{dagum1998openmp}) to enable multithreaded parallelism. Additionally, the Process Management Interface \texttt{pmi} is employed for managing the startup and communication between processes (e.g., in MPI~\cite{walker1996mpi}).

\section{Related Work} \label{section:related-work}
We discuss related data collection frameworks here with a focus on software tracking that includes process-level metadata, environment information, and executable file metadata collection.

XALT~\cite{agrawal2014user} is an infrastructure designed to systematically capture and analyze the software environment of HPC jobs. 
Building on earlier prototypes like ALTD~\cite{fahey2010automatic}, XALT employs LD\_PRELOAD techniques to inject data collection code into job processes, collecting information such as compiler usage, linked libraries, and environment variables. 
XALT creates \texttt{.json} files for every hooked process;
these files are periodically collected and consolidated into a dedicated database.
The information stored in the database can be used to generate job analytics, identify configuration issues, and guide software support and resource management decisions. 
XALT collects process metadata, a \texttt{sha1} hash of the executable, modules, libraries, and imported packages (e.g. Python), as well as adding an optional watermark to executables built under XALT for further tracking.

The Lightweight Distributed Metric Service (LDMS)~\cite{agelastos2014lightweight} is a low overhead, scalable infrastructure designed for the continuous monitoring of large-scale HPC systems and applications by allowing the collection of hundreds of metrics, from CPU, memory and network data to file system data.
Its modular design, which includes sampler, aggregator, and storage components, allows dynamic configuration, efficient data aggregation, and minimal performance impact even when deployed across thousands of nodes, demonstrated on platforms like NCSA's Blue Waters.
LDMS has recently been shown to be able to collect process-level data (from \texttt{/proc/\$pid/})~\cite{shoga2024evolving}, integrating system-level and application-specific data.

Similarly to XALT and LDMS, SIREN also collects many of the same fields,
such as process-level metadata, executable file metadata, loaded modules, shared libraries, and memory-mapped files.
In addition, SIREN incorporates digital forensic techniques in the form of fuzzy hashing of different characteristics of executables (including hashing of the raw binary content, printable strings, and symbol tables), which is used in cybersecurity scenarios for malware detection.
Moreover, with its UDP-based messaging and data collection, SIREN follows a design principle that focuses on \textit{graceful failure} when faced with errors.
The use of UDP data transmission (instead of TCP or creating individual files for every hooked process) minimizes the impact on user processes and file system by avoiding potential slowdown or failure points, such as socket exhaustion, excessive open file handles, and/or aggregating excessive amounts of small files.

\section{Conclusion and Future Work} \label{section:conclusion}
This work introduced SIREN, a data collection framework designed to gain insight into software usage in HPC systems by enabling analysis of process-level information, such as executable file metadata, compilers, shared libraries, imported Python packages, and fuzzy hashes of the application executables themselves.
SIREN combines data collection through the LD\_PRELOAD-based process hooking with fuzzy hashing through SSDeep, an approach inspired by malware detection for cybersecurity, to enable the identification and recognition of HPC software.

SIREN is deployed as an opt-in module on LUMI. 
The data collected, anonymized, analyzed, and presented here from voluntary LUMI users do not represent the broader software usage behavior on LUMI.
The analysis presented here shows how information about compilers and shared libraries can give insight into software usage on HPC systems and how system operators can use fuzzy hashes of executables to identify unknown application instances.
The collected data can be used not only to report software usage statistics, but also to help user support teams detect and mitigate problems regarding unexpected software usage and inefficiencies, as well as to troubleshoot user tickets through access to additional (and often hard to collect) information about the executed applications and their environment.

\textbf{Future Work.}
In the future, we plan to extend SIREN's deployment and data collection on LUMI to a broader cohort, as well as extract information from processes inside containers to analyze software usage at scale.
We also plan to cross-reference Python imports against known non-secure packages to detect known and potential vulnerabilities. \\

\noindent \textbf{Acknowledgments.}
The authors acknowledge the CSC IT Center for Science, Finland, for granting this project access to the EuroHPC LUMI supercomputer, for computational resources, and for generous support from the administrators. 
We thank all participants in the first opt-in data collection deployment campaign on LUMI for providing data.
The authors also thank HPE for their advice and support of this work through the HPC/AI EMEA Research Lab.

\bibliographystyle{splncs04}
\bibliography{main}

\end{document}